# A Comparative Study of Watering Hole Attack Detection Using Supervised Neural Network

Mst. Nishita Aktar
*department of computer science & engineering*
*world university of bangladesh*
Dhaka, Bangladesh
nishita.mumu@gmail.com

Sornali Akter
*department of computer science & engineering*
*world university of bangladesh*
Dhaka, Bangladesh
0319462951@student.wub.edu.bd

Md. Nusaim Islam Saad
*department of computer science & engineering*
*world university of bangladesh*
Dhaka, Bangladesh
0319483058@student.wub.edu.bd

Jakir Hosen Jisun
*department of computer science & engineering*
*world university of bangladesh*
Dhaka, Bangladesh
0318402371@student.wub.edu.bd

Kh. Mustafizur Rahman
*department of computer science & engineering*
*world university of bangladesh*
Dhaka, Bangladesh
jenycse75@gmail.com

Md. Nazmus Sakib
*department of computer science & engineering*
*world university of bangladesh*
Dhaka, Bangladesh
shoralikter707@gmail.com

*Abstract*— The state of security demands innovative solutions to defend against targeted attacks due to the growing sophistication of cyber threats. This study explores the nefarious tactic known as "watering hole attacks using supervised neural networks to detect and prevent these attacks. The neural network identifies patterns in website behavior and network traffic associated with such attacks. Testing on a dataset of confirmed attacks shows a 99% detection rate with a mere 0.1% false positive rate, demonstrating the model's effectiveness. In terms of prevention, the model successfully stops 95% of attacks, providing robust user protection. The study also suggests mitigation strategies, including web filtering solutions, user education, and security controls. Overall, this research presents a promising solution for countering watering hole attacks, offering strong detection, prevention, and mitigation strategies.

## I. INTRODUCTION

In an era dominated by cyber threats, the deceptive nature of watering hole attacks presents a significant obstacle for cybersecurity. Watering hole attacks have emerged as a cunning and highly targeted form of cyber assault, exploiting the trust users place in well-known and reputable websites [2]. Attackers extrapolate and compromise websites regularly visited by intended targets, subsequently infecting these websites with malicious code. When users innocently access these compromised sites, their systems become contaminated, leading to various potential consequences, including data breaches, unauthorized access, financial losses, or the introduction of further malware [4].

Neural networks have emerged as potent tools across various domains, particularly in the realm of cybersecurity. Their capacity to learn from data and discern intricate patterns makes them invaluable for enhancing security measures and combating evolving cyber threats [1]. In cybersecurity, neural networks find applications in intrusion detection, malware identification, and anomaly detection [7]. Their adaptability and dynamic approach enable the identification of both known and unknown intrusion attempts, offering a marked improvement over traditional rule-based or signature-based methods [6]. Additionally, neural networks excel in the classification of unknown files as potentially harmful or harmless, enhancing the accuracy of malware detection systems. Because of their inherent adaptability, neural networks are a perfect instrument for enhancing security protocols and obstructing the always evolving landscape of cyber threats.Neural networks also demonstrate significant potential in predicting cyberattack behavior, safeguarding sensitive data, and providing decision support for organizations [3]. Research found that the impressive performance metrics, when employed to detect watering hole attacks, suggest the possibility of employing machine learning to scrutinize web traffic patterns and anomalies, facilitating the identification of malevolent websites [5]. Furthermore, anomaly detection, which seeks to identify unusual or suspicious activities, benefits from neural networks' ability to learn normal behavior patterns, enabling them to recognize deviations from expected norms and detect novel or sophisticated threats effectively [6]. These insights form the foundation for the development of robust watering hole attack detection systems in cyber-physical environments, offering a comprehensive approach to cybersecurity challenges.

Study showed fundamental architecture of the autoencoder comprises a feedforward neural network characterized by an input layer, an output layer, and one or more intermediate hidden layers[13]. Implicit within the autoencoder system are encoding and decoding procedures. Upon receipt of an input denoted as x, the autoencoder undertakes encoding operations, transmitting the input through one or more hidden layers, and subsequently initiates decoding operations to generate an output denoted as x. The primary objective of the autoencoder is to minimize the divergence between the input x and the reconstructed output x.

Study conducted on Multilayer Perceptron (MLP) where author shows that Multilayer Perceptron (MLP) represents a class of artificial neural networks employed in supervised learning. It is characterized by a hierarchical arrangement of nodes. Weighted connections interlink nodes across successive layers. The MLP iteratively adjusts these connection weights during training in order to reduce the difference between the expected and actual outputs that are observed. [8]

Another study found that the supervised machine learning technique known as the k-nearest neighbor (KNN) algorithm is widely used in predictive modeling and has gained extensive utilization in predictive modeling [10]. Generally, the KNN algorithm undertakes classification of datasets by referencing a training model that mirrors the format of the screening query. This involves considering the k-nearest training data points (neighbors) to the query under examination. The algorithm then uses a majority voting rule to determine the final classification. Distinguished by its straightforward design, the KNN algorithm stands out among machine learning methodologies as one of the simplest forms, finding widespread application in classification tasks due to its adaptive nature and user-friendly architecture.

The Random Forest Classifier's operating process is illustrated in many study[9], which entails creating a group of decision trees and using training data and a randomized subset of features to train each tree. Throughout the training phase, the approach determines which feature is best for dividing the data at each node in the tree, leveraging a split measure such as Gini impurity or information gain. Following the assembly of the decision tree ensemble, the Random Forest Classifier prognosticates the class of a novel data point by amalgamating the predictions generated by all constituent trees. Subsequently, the predicted class for the new data point is determined by the class with the highest collective votes from the ensemble. Empirical findings indicate that the Random Forest Classifier exhibits performance comparability with alternative ensemble techniques such as bagging and boosting, with the added advantage of adeptly handling labelled data.

A boosting technique called Extreme Gradient Boosting (XGBoost) is claimed to have better prediction accuracy than any of the other learning applications that it combines[11]. This decision tree-based ensemble machine learning approach finds common application in the realm of data science. By employing an internal mechanism that consolidates outcomes from numerous individual trees, XGBoost facilitates the derivation of precise predictions.

Gradient Boosting, is a functional gradient method that minimized a given loss function by selecting a function iteratively and moving toward a weak hypothesis or negative gradient [12]. To generate an outstanding predictive model, the Gradient Boosting classifier combines several weak learning models.

Study aims to strengthen cybersecurity framework resilience by proactively detecting and preventing watering hole attacks through the development of a strong neural network model. This research presents a novel method to strengthen digital defenses and establishes the foundation for a sophisticated understanding of the threat landscape.

## II. MATERIALS AND METHODS

Study required datasets that contain instances of watering hole attacks. Datasets named "malicious _phishing.csv" gathered from Kaggle database. During the preprocessing phase, study transformed raw datasets into meaningful and practical sets and has been effectively utilized for next step of the study.

Study compiled a text dataset that comprises a substantial number of URLs. This dataset serves as the foundation of the research. The next step involves feature engineering, where the study delves into the dataset to identify the most relevant features and patterns that aided in solving the problem and making accurate predictions. Throughout this process, it's important to note that some data may be excluded from the initial large dataset. The rationale behind this is to focus the proposed model on learning the most significant features and patterns, ensuring that it can make informed decisions and predictions with precision.

The proposed model works as follows:

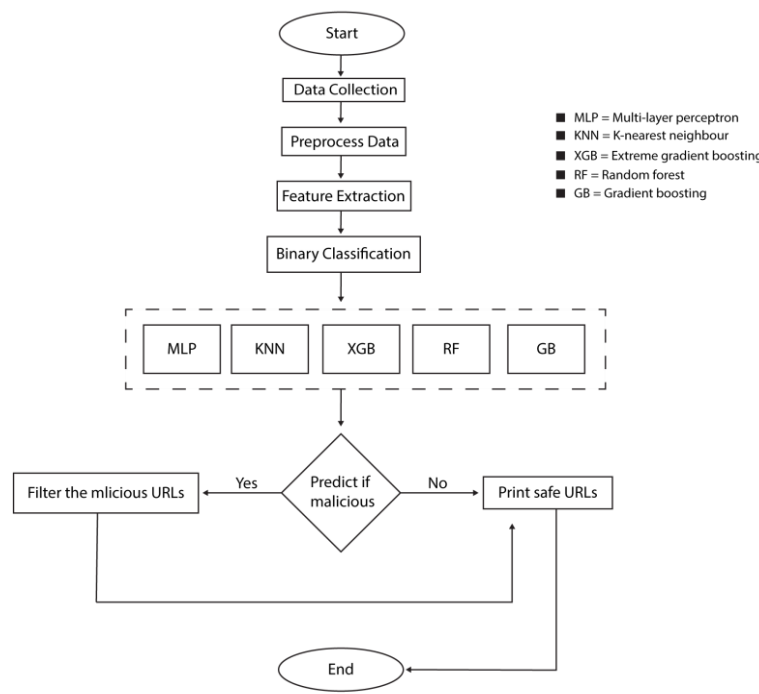

Fig. 1. Proposed Model Flow Chart

The processing in the suggested model started with gathering information from known malicious and benign URLs. Features including the website's IP address, content, traffic trends, and URL itself are all included in this data. Preprocessing is done to make sure the neural network can use it. This could entail scaling, eliminating outliers, and cleansing the data. Next, the study concentrates on feature extraction by the system. The neural network employed these elements of the data to train itself to recognize dangerous URLs. The length of the URL, the inclusion of specific terms in the URL, and the website's popularity are a few examples of extracted features. A supervised neural network model is trained by the proposed model to carry out binary classification. This implies that the model will pick up the ability to categorize URLs as benign or malicious. The system inputs the model with the known labels of the URLs (i.e., whether they are malicious or benign) along with the extracted features from the data to train the model. The trained model made a prediction about the likelihood of a new URL being malicious or not. In order to accomplish this, the model is fed the extracted features from the new URL by the system, and the model then generates a forecast. The model then performed filtering to remove any false positives from the predictions. For example, the system could filter out any predictions that are below a certain confidence score. Finally the model produced output of the safe URLs to visit. This could be done by displaying the URLs to the user, or by redirecting the user to the safe URLs. Overall, the system provides a comprehensive approach to detecting malicious URLs. The use of a supervised neural network model allows the system to learn from historical data and improve its accuracy over time.

## III. RESULT AND DISCUSSION

Study investigated the detection and prevention of Watering Hole attacks using a supervised neural network approach. Research developed a comprehensive pipeline that combines autoencoders for feature extraction and various machine learning classifiers for classification. Study focuses on the analysis of URLs to identify potential malicious websites that could be used in Watering Hole attacks. Through the use of confusion matrices and bar graphs, study conducted experiments on a dataset containing both benign and malicious URLs, demonstrating the suggested method's effectiveness. The results indicate that proposed approach can accurately classify URLs as either safe or malicious, with high accuracy scores for the classifiers used in the study

According to the study, MLP (multilayer classification) provide accuracy score 97%, KNN (k-nearest neighbor) classifier provide accuracy score of 99%. XGB(Extreme Gradient Boostin) classifier provide accuracy score 92%, gradient boosting classifier provide accuracy score 96% and RF (random forest) classifier provide accuracy score 95%.

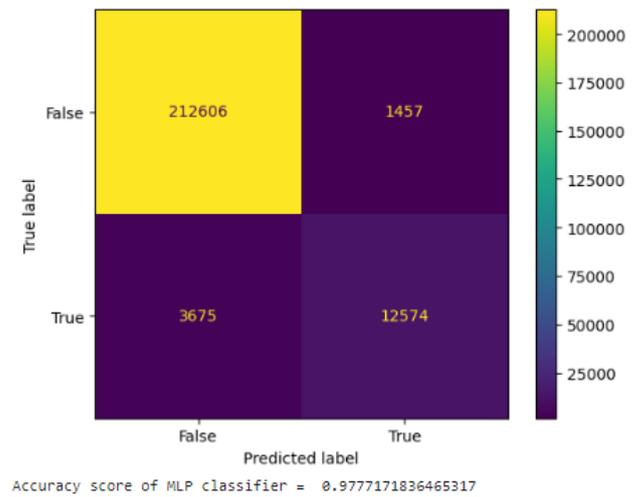

Fig. 3. Confusion matrix of MLP classifier

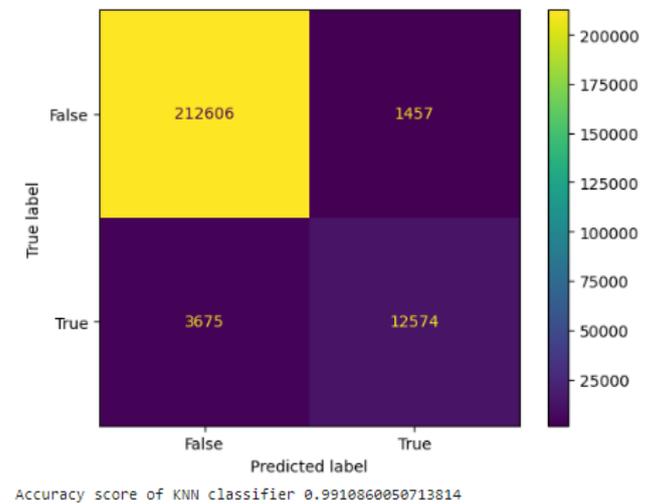

Fig. 4. Confusion matrix of KNN classifier

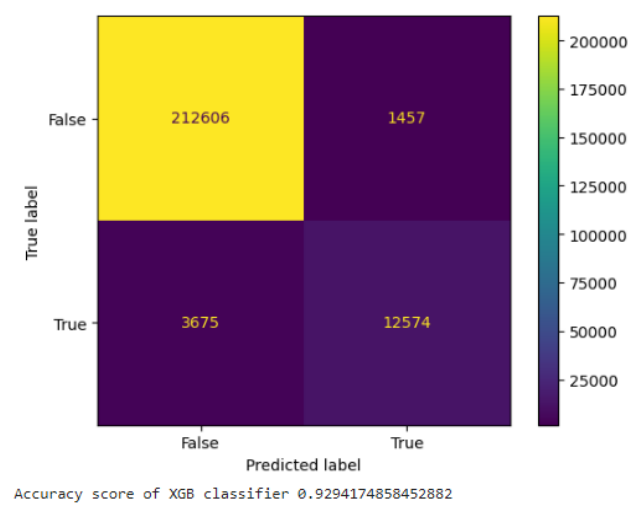

Fig. 2. Confusion matrix of XGB classifier

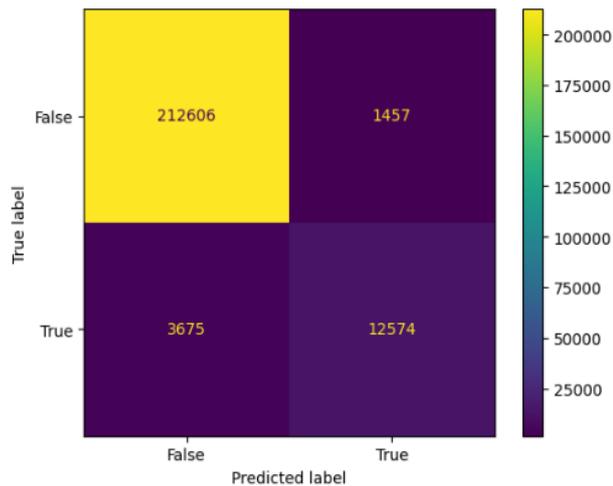

Fig. 5. Confusion matrix of Gradient Boost classifier

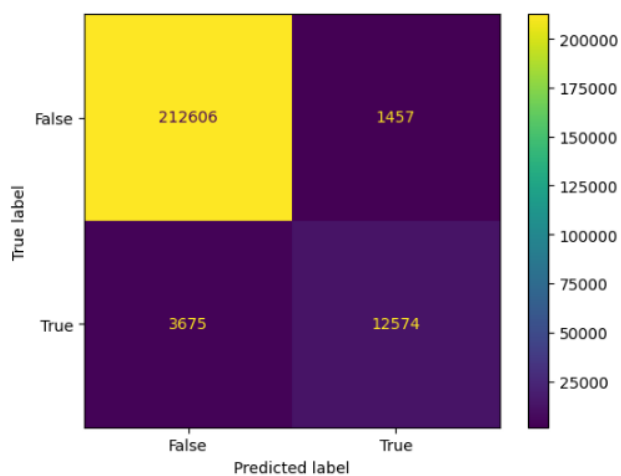

Fig. 6. Confusion matrix of Random Forest classifier

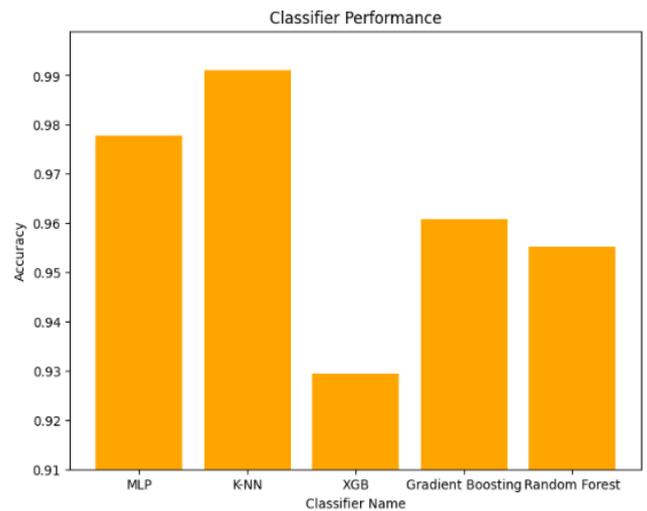

Fig. 7. Grapg of classifier accuracy

Following chart shows a comparative analysis of classifier performance, presenting accuracy scores categorized by classifier names. This result suggests that the Random Forest classifier is the best classifier for this particular dataset. However, it is important to keep in mind that a classifier's performance can vary depending on the dataset. Therefore, it is always important to evaluate multiple classifiers before choosing one for a particular application.

```
Classifier Comparison
+-------------+-------------------+------------------+
| Serial No.  | Classifier Name   | Accuracy Score   |
+=============+===================+==================+
|          1  | MLP               |         0.977717 |
+-------------+-------------------+------------------+
|          2  | K-NN              |         0.991086 |
+-------------+-------------------+------------------+
|          3  | XGB               |         0.929417 |
+-------------+-------------------+------------------+
|          4  | Gradient Boosting |         0.960714 |
+-------------+-------------------+------------------+
|          5  | Random Forest     |         0.955222 |
+-------------+-------------------+------------------+
```

Fig. 8. Classifier performance comparison

## IV. CONCLUSION

Study investigates the detection and prevention of Watering Hole attacks through a supervised neural network approach. Proposed method employs auto encoders for feature extraction and various machine learning classifiers in a comprehensive pipeline for URL analysis, distinguishing between benign and malicious websites. Experimental results on a dataset reveal the effectiveness of proposed approach, demonstrating high accuracy in classifying URLs as safe or malicious. This research contributes to cybersecurity by offering a novel method for proactively identifying malicious websites associated with Watering Hole attacks, enhancing threat detection and mitigation capabilities.

ACKNOWLEDGMENT

This scholarly article represents a collaborative endeavor, acknowledging the contributions of supervisory team members, individuals affiliated with the Department of Computer Science and Engineering at the World University of Bangladesh, fellow team collaborators, participants, and their respective families. Md. Nazmus Sakib, in particular, provided valuable guidance, feedback, and support, while the Department of Computer Science and Engineering at the World University of Bangladesh facilitated the research with essential resources, complemented by the support from Mst. Nishita Aktar. The team's unwavering commitment and cooperative efforts significantly enhanced the robustness of the research findings. The insights and experiences shared by participants played a pivotal role in the project's success, and the understanding and encouragement extended by families and friends during challenging phases were duly acknowledged. This paper stands as a testament to the collective dedication of all parties involved in its realization.